\documentclass{aastex63}
\usepackage{xcolor}
\usepackage{longtable}
\usepackage{amsmath,amssymb}
\usepackage{rotating}
\usepackage{hyperref}
\usepackage{graphicx}

\submitjournal{ApJL}

\shorttitle{The firm detection of the highest energy of fundamental cyclotron lines}
\shortauthors{Ge et al.}

\begin{document}

\title{ \textit{Insight-HXMT} firm detection of the highest energy fundamental cyclotron resonance scattering feature in the spectrum of GRO J1008-57}

\author{M. Y. Ge}
\affil{Key Laboratory for Particle Astrophysics, Institute of High Energy Physics, Chinese Academy of Sciences, 19B Yuquan Road, Beijing 100049, China}

\author{L. Ji}
\email{ji.long@astro.uni-tuebingen.de}
\affil{Institut f{\"u}r Astronomie und Astrophysik, Kepler Center for Astro and Particle Physics, Eberhard Karls, Universit{\"a}t, Sand 1, D-72076 T{\"u}bingen, Germany}

\author{S. N. Zhang}
\email{zhangsn@ihep.ac.cn}
\affil{Key Laboratory for Particle Astrophysics, Institute of High Energy Physics, Chinese Academy of Sciences, 19B Yuquan Road, Beijing 100049, China}
\affil{University of Chinese Academy of Sciences, Chinese Academy of Sciences, Beijing 100049, China}

\author{A. Santangelo}
\affil{Institut f{\"u}r Astronomie und Astrophysik, Kepler Center for Astro and Particle Physics, Eberhard Karls, Universit{\"a}t, Sand 1, D-72076 T{\"u}bingen, Germany}

\author{C. Z. Liu}
\affil{Key Laboratory for Particle Astrophysics, Institute of High Energy Physics, Chinese Academy of Sciences, 19B Yuquan Road, Beijing 100049, China}

\author{V. Doroshenko}
\affil{Institut f{\"u}r Astronomie und Astrophysik, Kepler Center for Astro and Particle Physics, Eberhard Karls, Universit{\"a}t, Sand 1, D-72076 T{\"u}bingen, Germany}

\author{R. Staubert}
\affil{Institut f{\"u}r Astronomie und Astrophysik, Kepler Center for Astro and Particle Physics, Eberhard Karls, Universit{\"a}t, Sand 1, D-72076 T{\"u}bingen, Germany}

\author{J.~L. Qu}
\affil{Key Laboratory for Particle Astrophysics, Institute of High Energy Physics, Chinese Academy of Sciences, 19B Yuquan Road, Beijing 100049, China}

\author{S. Zhang}
\affil{Key Laboratory for Particle Astrophysics, Institute of High Energy Physics, Chinese Academy of Sciences, 19B Yuquan Road, Beijing 100049, China}

\author{F. J. Lu}
\affil{Key Laboratory for Particle Astrophysics, Institute of High Energy Physics, Chinese Academy of Sciences, 19B Yuquan Road, Beijing 100049, China}

% ...




\begin{abstract}
We report on the observation of the accreting pulsar GRO J1008-57 performed by \textit{Insight-HXMT} at the peak of the source's 2017 outburst. Pulsations are detected with a spin period of 93.283(1)\,s. The pulse profile shows double peaks at soft X-rays, and only one peak above 20\,keV. The spectrum is well described by the phenomenological models of X-ray pulsars. A cyclotron resonant scattering feature is detected with very high statistical significance at a centroid energy of $E_{\rm cyc}=90.32_{-0.28}^{+0.32}$\,keV, for the reference continuum and line models, HIGHECUT and GABS respectively. Detection is very robust with respect to different continuum models. The line energy is significantly higher than what suggested from previous observations, which provided very marginal evidence for the line. This establishes a new record for the centroid energy of a fundamental cyclotron resonant scattering feature observed in accreting pulsars. We also discuss the accretion regime of the source during the \textit{Insight-HXMT} observation. 

\end{abstract}



\section{Introduction} 
Accreting X-ray pulsars are neutron stars in binary systems, in which the strongly magnetised neutron star accretes matter from a donor star (typically a young O or B star).
The magnetic fields of the neutron stars are thought to have strengths of $B\sim 10^{12}$\,G.
One of the most solid probes of the magnetic fields of accreting pulsars is the observation of cyclotron resonant scattering features (CRSFs). 
These scattering features appear as broad absorption lines observed in the source's spectrum at hard X-ray energies corresponding to transitions between the discrete Landau levels of electrons' motion perpendicular to the magnetic field lines. 
The first cyclotron line was discovered in 1976 in a balloon observation of Her X-1 \citep{1977NYASA.302..538T} and correctly interpreted by \citet{Truemper1978}.
Since then, CRSFs have been found in several tens of sources \cite[for a recent review, see][]{Staubert2019}. Several systems exhibit harmonics in addition to the fundamental \citep[see, e.g., ][]{Santangelo1999}.
The centroid energy of the fundamental and harmonic CRSF lines is given, in the non-relativistic approximation, by $E_{\rm cyc}\sim11.6\,nB_{12}(1+z)^{-1}$\,keV, where $z$ is the gravitational redshift, $B_{12}$ is the magnetic field in the line forming region in units of $10^{12}$\,G, and $n$ is the quantum number.
In most X-ray pulsars, CRSF energies are observed to vary with the pulse phase, which is explained with the change of the view angle with respect to the line-forming region.
In addition, the line energy has been observed to vary with the luminosity. A 'positive' correlation between the line centroid and the luminosity is observed for low luminosity sources. On the other hand, an anti-correlation, i.e., 'negative' correlation, has been observed for high luminosity sources \citep[see, e.g., ][]{Staubert2007, Doroshenko2017, Vybornov2018}. This is explained invoking different accretion regimes in the so-called sub-/super critical accretion \citep[for details, see, e.g., ][]{Basko1976,Becker2012, Mushtukov2015}.

GRO J1008-57 was discovered with the \textit{Compton Gamma Ray Observatory (CGRO)}/BATSE during a bright outburst in 1993 \citep{Stollberg1993}. The distance to the source has been estimated to be 5.8$\pm$0.5\,kpc \citep{Riquelme2012}. GRO J1008-57 is a transient high mass X-ray binary (HMXB) with a Be companion.
It exhibits periodic Type-I outbursts around the periastron passage, and sometimes irregular Type-II outbursts, which are brighter and longer.
\citet{Bellm2014, Yamamoto2014} reported the marginal detection at $\sim$ 4\,$\sigma$ of a CRSF line around 78\,keV, by using non-simultaneous observations of \textit{Swift}, \textit{Suzaku} and \textit{NuSTAR} taken during the 2012 outburst of the source.
Other observations of different outbursts with CGRO and INTEGRAL hinted at a similar result but at even lower significance \citep{Grove1995, Shrader1999, Kuehnel2013, Wang2014}.
In this paper, we report the highly significant
detection of a CRSF in the spectrum of GRO J1008-57 at $E\sim$90\,keV. This is the highest energy fundamental CRSF ever observed. The line was observed during the outburst of GRO J1008-57 in 2017 with \textit{Insight-HXMT}.

\section{observations and results}
The source was observed with \textit{Insight-HXMT} for 50\,ks at the peak of a Type-II outburst on 12th August 2017 (MJD 57977; see Figure 1, left panel).
\textit{Insight-HXMT} is the first Chinese X-ray astronomy satellite, launched on June 15, 2017 \citep{Zhang2020}.
The instruments of the scientific payload include the high energy detectors (HE) \citep{Liu2020}, the medium energy detectors (ME) \citep{Cao2020}, and the low energy detectors (LE) \citep{Chen2020}, which have an effective area of 5100\,$\rm {cm}^2$, 952\,$\rm {cm}^2$ and 384\,$\rm {cm}^2$, respectively.
We performed our analysis by using the {\sc hxmtsoft} \footnote{ \url{http://www.hxmt.org/index.php/dataan}} analysis package and the official user guides \footnote{\url{http://www.hxmt.org/sjfxwd/65.jhtml}}.
The background was estimated and subtracted using the standalone python scripts {\sc lebkgmap}, {\sc mebkgmap} and {\sc hebkgmap} \citep{Li2020, Guo2020, Liao2020, Liao2020b}. 
We screened the data according to the suggested criteria of the good-time-interval (GTI) selection: an elevation angle (ELV) larger than 10$^\circ$; geometric cutoff rigidity (COR) larger than 8\,GeV; offset for the point position smaller than 0.1$^\circ$; at least 300\,s before and after the South Atlantic Anomaly passage. 

In the spectral analysis, we used the well-calibrated energy bands of LE, ME and HE: 2-10\,keV, 8-30\,keV and 30-135\,keV, respectively.
We searched for pulsations by using the $Z^2_{2}$-test with a step of the Fourier frequency \citep{Buccheri1983}, and by taking into account barycentric and orbital motion correction. 
We found that the source pulsates with a spin period of 93.283(1)\,s.

\begin{figure}
	\centering
	\includegraphics[width=0.5\textwidth, height=4.3in]{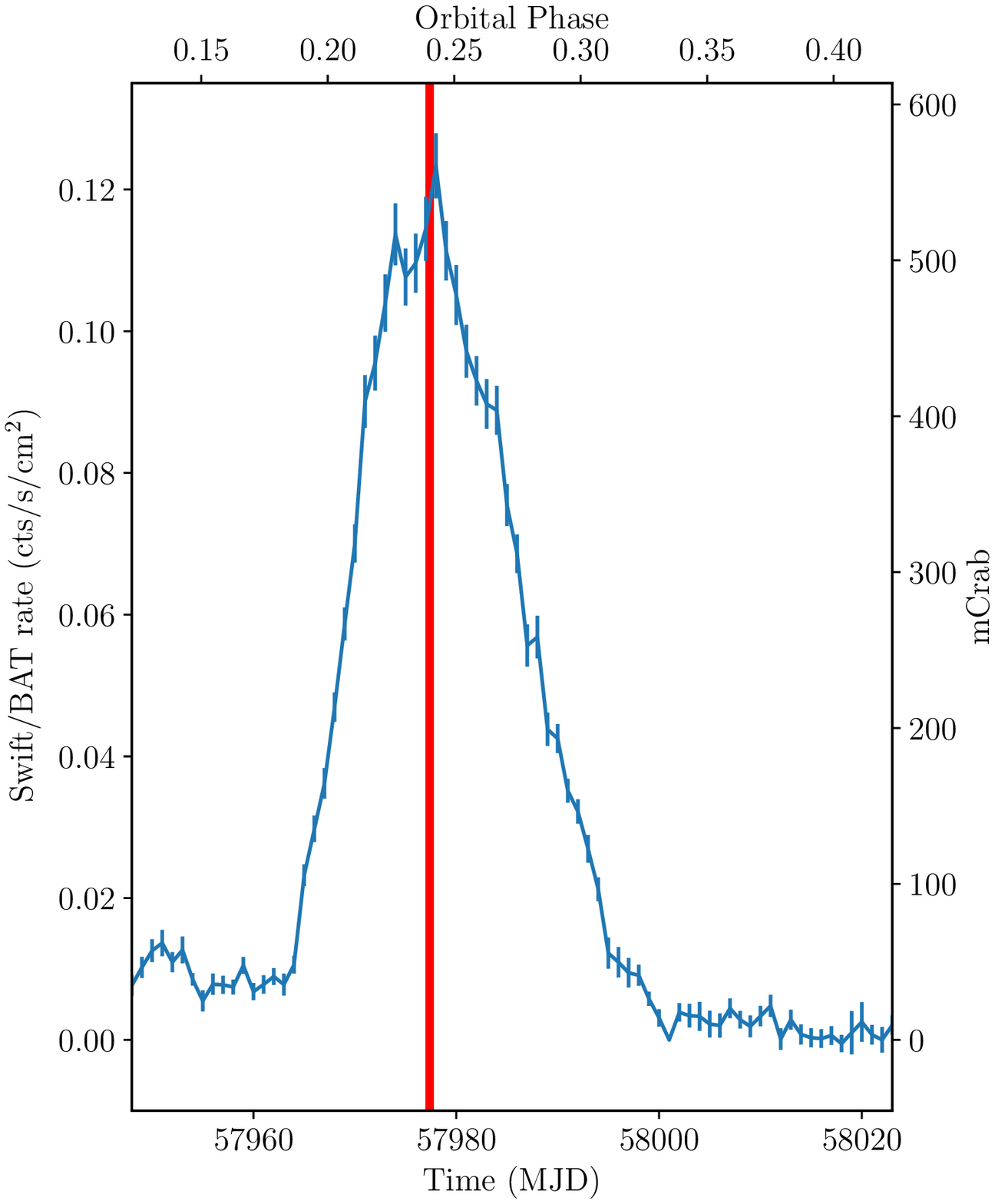} 
	\includegraphics[width=0.45\textwidth]{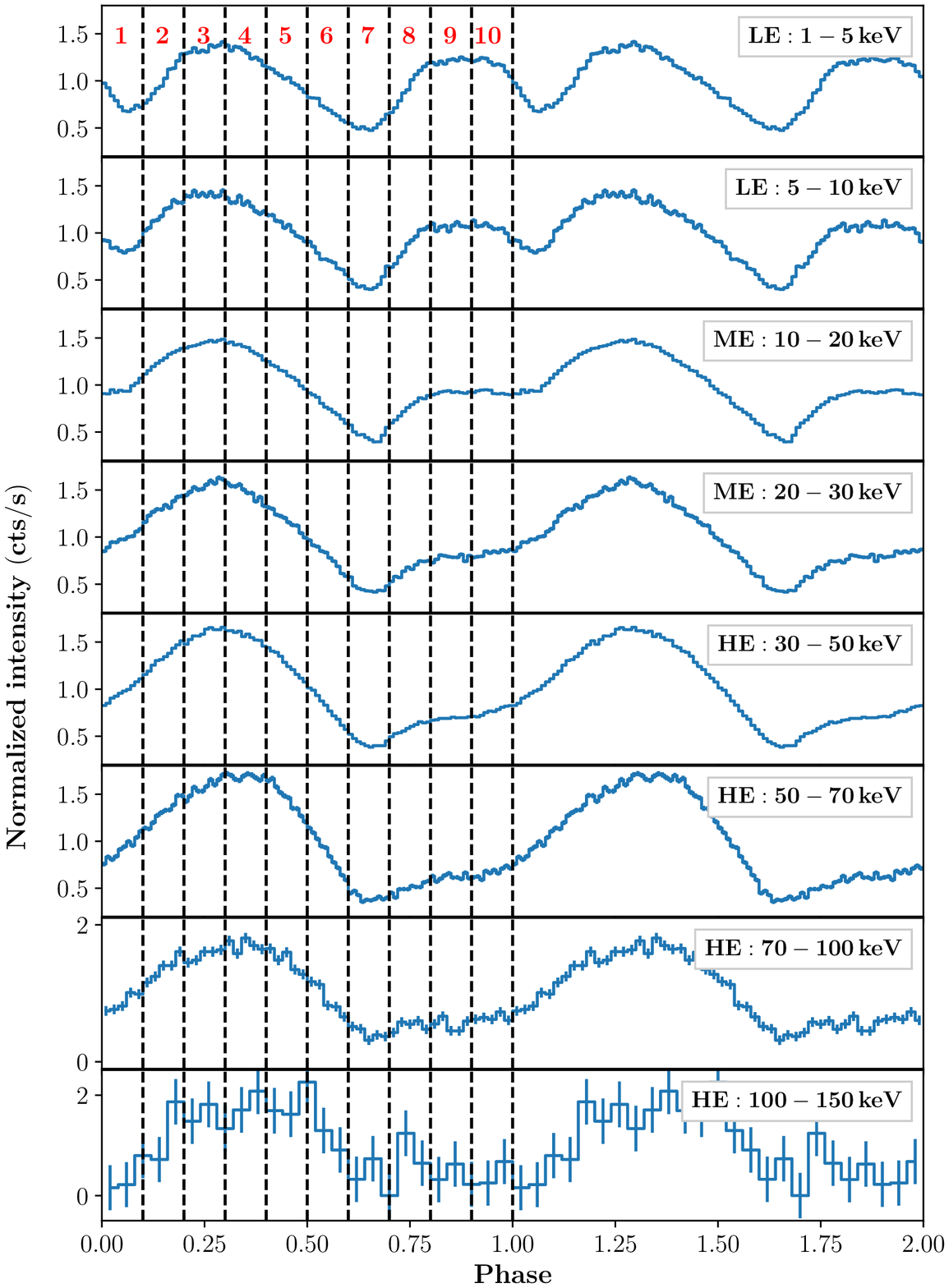}
	\caption{  
	Left panel: The \textit{Swift}/BAT lightcurve of GRO J1008-57 during its outburst in 2017. The red line represents the observational time of \textit{Insight-HXMT}. The orbital phase is calculated based on the previously reported ephemeris \cite{Kuehnel2013}.
	Right panel:
	The pulse profiles in different energy bands of GRO J1008-57 around the outburst peak in 2017. The spin period is 93.283(1)\,s.
	}  
	\label{pulse}  
\end{figure}

After folding the events, we obtained pulse profiles that clearly depend on the energy (shown in Figure~\ref{pulse}, right panel).
Similar as in other pulsars, the shape of the pulse profile is more complex in the low energy band, exhibiting double peaks, and gradually evolves into a single peak profile at higher energies.

For the spectral analysis, we fitted the wide-band phase-averaged spectrum with several phenomenological models typically used to model spectra of accreting pulsars \citep{Staubert2019}. We show here results from two of the models: the HIGHECUT model and the NPEX model \citep{Mihara1995}. 

\begin{figure}
	\includegraphics[width=0.49\textwidth]{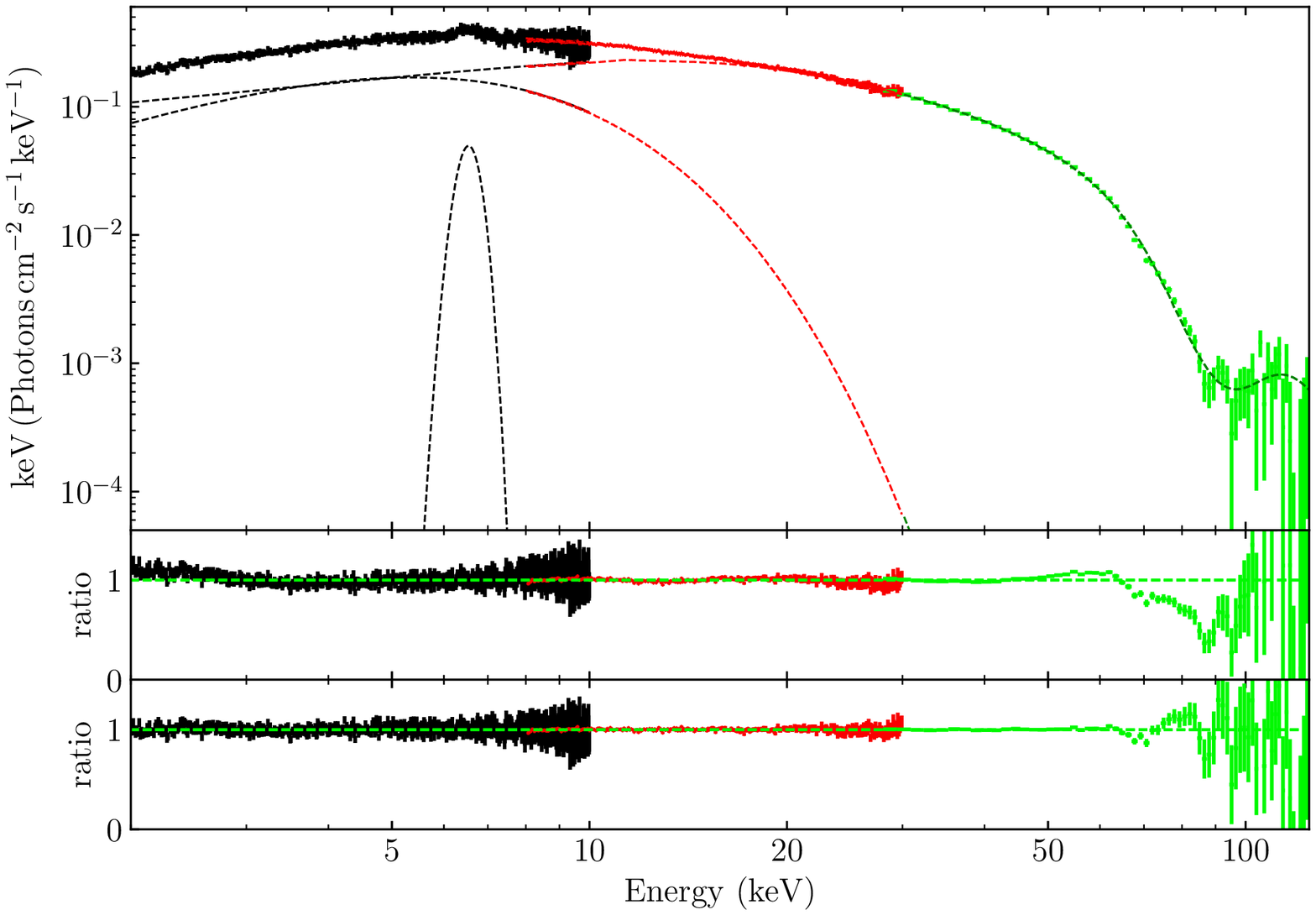}
	\includegraphics[width =0.49\textwidth]{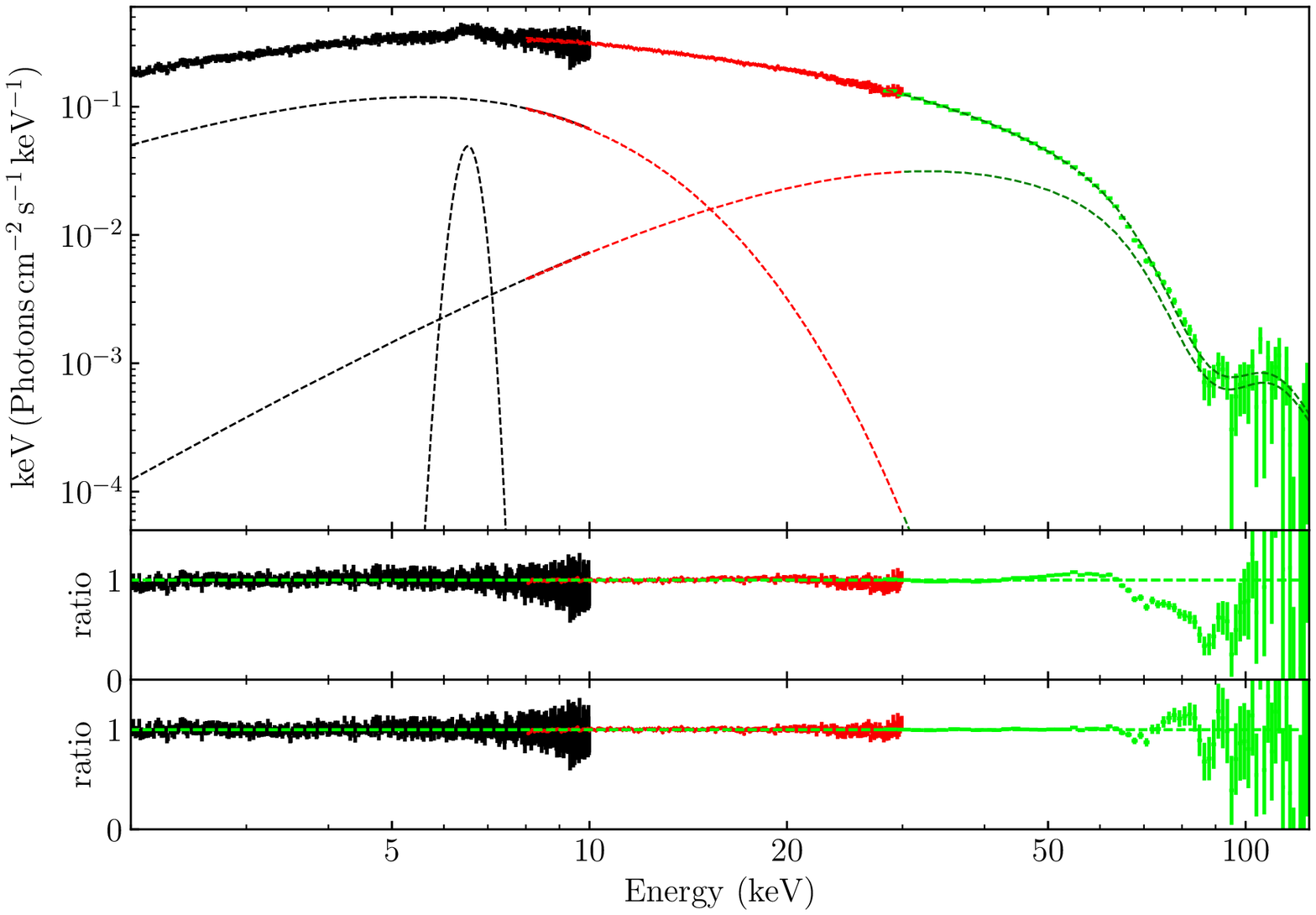}
	\caption{ 
		The unfolded phase-averaged spectrum of GRO J1008-57 observed in 2017 outburst fitted by HIGHECUT (left panel) and NPEX (right panel) models. The black, red and green lines represent LE, ME and HE energy ranges, respectively.  
		We also show ratios, i.e., the data divided by the folded model, with and without the CRSF in the bottom and middle panels. 
	}  
	\label{fitting}  
	
\end{figure}

The former is a product of a power-law model and a multiplicative exponential factor \citep{Staubert2007}:
$$ I_{\rm E}=\left\{
\begin{aligned}
& K \cdot E^{-\Gamma},\ if\ E \leq E_{\rm cut}\\
& K  \cdot E^{-\Gamma} {\rm exp}(-\frac{E - E_{\rm cut}}{E_{\rm f}}),\ if\ E \geq E_{\rm cut}
\end{aligned}
\right.
$$
Since the function contains a discontinuity of its first derivative, we included a smooth function, i.e., a weak Gaussian absorption line with the centre energy fixed to $E_{\rm cut}$ \citep[for details, see, e.g., ][]{Coburn2002}. The HIGHECUT model is widely used in the literature, and chosen as reference in the review article by \citet{Staubert2019} to model all spectra of the accreting pulsars with CRSFs. 
The NPEX model is a sum of positive and negative cutoff power-law spectral models, which resembles a Comptonization spectrum \citep{Mihara1995}:
$$ I_{\rm E}=
\begin{aligned}
(K_{1} \cdot E^{-\Gamma_1} + K_{2} \cdot E^{+\Gamma_2} ){\rm exp}(-E/E_{\rm f})
\end{aligned}
$$
where $\Gamma_2$ is normally fixed to 2 \citep{Mihara1995}. The NPEX model is also largely used in literature, and for GRO J1008-57 allows a direct comparison of our findings with those of \cite{Yamamoto2014}.
Following earlier publications \citep{Kuehnel2013,Yamamoto2014, Bellm2014}, a blackbody component at soft X-rays was included.
In addition, we considered a Gaussian emission line around 6.4\,keV to account for the $K_{\rm \alpha}$ fluorescence iron line.
All uncertainties quoted in this paper correspond to a 68\% confidence level.
Fits with the above defined continuum models were 
statistically not acceptable since they resulted in a reduced-$\chi^2$ $>$ 2. In particular, significant residuals remained at hard X-rays ($>$\,50keV) (Figure~\ref{fitting}).

We found that an additional absorption-like CRSF component, e.g., GABS or CYCLABS in {\sc xspec}, improved improved the fits significantly, reaching a line significance in excess of 70 standard deviations (as found through Monte Carlo simulations {\sc SIMFTEST}.
\footnote{We simulated 1000 spectra using the best-fitting model where the CRSF is excluded, and then fitted these simulated spectra using the same model with and without the absorption. A distribution of $\Delta \chi^2$ can be obtained assuming that the CRSF is caused by statistical fluctuations. Thus the significance of the CRSF can be estimated by comparing the real observation with the simulated distribution.}
The aforementioned GABS and CYCLABS models are widely used in the literature to describe CRSFs \citep{Staubert2019, Mihara1990, Makishima1990}. In particular the GABS model is the reference model used in the review by \cite{Staubert2019}, while CYCLABS has been extensively used to model the CRSF in previous observations of GRO J1008-57. The two models are expressed as:
\begin{equation} 
 {\rm GABS}(E)={\rm exp}\{-\frac{D_{\rm cyc}}{\sqrt{2\pi}\sigma_{\rm cyc}} {\rm exp} \left[{-\frac{1}{2} (\frac{E-E_{\rm cyc}}{\sigma_{\rm cyc}})^2}\right] \}
\end{equation}
and 
\begin{equation} 
 {\rm CYCLABS}(E)={\rm exp}\{-D_{\rm f} \frac{(W_{\rm f}E/E_{\rm cyc})^2}{(E-E_{\rm cyc})^2+W_{\rm f}^2}\}
\end{equation}

We show the best fit spectra in Figure~\ref{fitting}, and best fit results in Table~\ref{pars}.
We observe that our findings on the line parameters do not significantly change when using other continuum models.

For completeness, we investigated the variability of the spectral shape with pulse phase.
We obtained spectra for ten phase bins (see Figure~\ref{pulse}), and analysed them using the aforementioned models.
For some of the pulse phases, the parameters of the CRSF are not well constrained because of poor statistics.
For these pulse phases, we have fixed the line width and depth at values obtained from the phase-averaged analysis.
As an example, we show the spectral parameters change with the phase of the pulse in Figure~\ref{phaseResolved}, for the HIGHECUT plus GABS model.
Both the CRSF and continuum parameters evolve with pulse phase significantly. The line energy varies with pulse phase up to $\sim$ 5\,keV (i.e, 6\%) and in general shows a lower value around the pulse peak.

\section{Discussion}

We report on the spectral analysis of the outburst of GRO J1008-57 observed with \textit{Insight-HXMT} in 2017. The spectral continuum is generally consistent with that reported in literature from observations of previous outbursts \citep{Kuehnel2013,Yamamoto2014, Bellm2014}.
In particular, we have observed with unprecedented high significance ($>$ 70\,$\sigma$) a CRSF at $E_{\rm cyc}=90.32_{-0.28}^{+0.32}$\,keV with a width ($\sigma_{\rm cyc}$) of $14.57_{-0.11}^{+0.14}$\,keV (combining the reference HIGHECUT and GABS models). 
The $E_{\rm cyc}$-$\sigma_{\rm cyc}$ relation is consistent with that of many other sources \citep[see, Fig. 7 in][]{Coburn2002}.

This $E_{\rm cyc}$ is the highest value for any fundamental CRSF reported so far. The source's magnetic field strength can be estimated to be $B=(1+z)\,E_{\rm cyc}/11.6=7.8\times 10^{12} (1+z)$\,G, where $z$ is the gravitational redshift at the line-forming region.

The line energy does not significantly depend on the choice of the continuum. However, it slightly changes (to $E_{\rm cyc}=83.00_{-0.63}^{+0.91}$) if the {\sc CYCLABS} model is used, together with NPEX. This is not surprising since the true minimum of the {\sc CYCLABS} line profile is lower by a factor of $1+(\sigma_{cyc}/E_{cyc})^2$ if compared to the simple gaussian profile. The line energy deduced with CYCLABS can thus be lower up to 20\%, compared to other models, especially when the line width is not well constrained, or actually frozen \citep{Mihara1995,Nakajima10,2015MNRAS.448.2175L}. 

The CRSF line energy observed by \textit{Insight-HXMT}, is higher compared to the marginal  (99\% confidence limit) detection at $E_{\rm cyc}=76_{-1.7}^{+1.9}$\,keV, reported by \citet{Yamamoto2014}, based on \textit{Suzaku} observations of the 2012 source's outburst. The discrepancy can be explained considering that \citet{Yamamoto2014} modelled the data with the CYCLABS and NPEX models, and by the limited quality of \textit{Suzaku} data at hard energies. Similar results at $\sim$ 4\,$\sigma$ were obtained by \citet{Bellm2014}, who modelled a combination of \textit{Suzaku} and \textit{NuSTAR} data with CYCLABS. The \textit{Insight-HXMT} detection is also significantly higher than the hint  of a CRSF reported by \cite{Kuehnel17} based on non-simultaneous \textit{NuSTAR-Swift}, and \textit{Suzaku} data at different epochs. Systematic uncertainties caused by the changes of the continuum spectrum were not taken into account, which might influence the line detection. 
Another source of discrepancy, might be related to the energy band used to model the broad-band spectrum, which is broader for our \textit{Insight-HXMT} observation. 
We note, however, that our result is in agreement with \textit{OSSE}'s observation of 1993 and the joint fit of the \textit{RXTE} and \textit{Suzaku} spectra \citep{Kuehnel2013}, although these detections have a very low significance level (less than 2\,$\sigma$).

The bolometric luminosity during the \textit{Insight-HXMT} observation in the energy range of 1-100\,keV is $L_{\rm X} \sim \rm 5.8\times10^{37}\,erg\,s^{-1}$, for a distance of 5.8\,kpc \citep{Riquelme2012}, lower than that observed in 2012, $L_{\rm X} \sim \rm 1.1\times10^{38}\,erg\,s^{-1}$, and comparable with observations reported by \citet{Kuehnel2013}.

On the base of our single observation, we cannot confirm or exclude whether the line centroid $E_{\rm cyc}$ varies with luminosity as reported by \citet{Yamamoto2014}, who suggested that the CRSF energy could be negatively correlated with the luminosity like in the case of the two other accreting pulsars V 0332+53 and SMC X-2 \citep{Tsygankov2006, Jaisawal2016}. If confirmed, the negative correlation would imply that the source is in the super-critical accretion regime \citep{Basko1976}, and the source luminosity must be higher than the critical luminosity $L_{\rm crit}$.

Unfortunately, the comparison of our \textit{Insight-HXMT} observation's luminosity with predictions for $L_{\rm crit}$ found in literature is also inconclusive. According to \citet{Becker2012}, $L_{\rm crit}$ is given by $\sim 1.5\times 10^{37}\,B_{12}^{16/15}$\,$\rm erg\,s^{-1}$, that for our estimate of the B field implies $L_{\rm crit} \sim 1.7 \times 10^{38} {\rm erg\,s^{-1}}$. The source would be therefore in the sub-Eddington accretion regime. On the other hand, predictions by \cite{Basko1976}, and more recently by \citet{Mushtukov2015} (see their Figure~7), suggest lower values of the critical luminosity ($\lesssim 5\times10^{37}$\,${\rm erg\,s^{-1}}$), comparable with the luminosity of our observation.

\begin{figure}
	\includegraphics[width=0.5\textwidth]{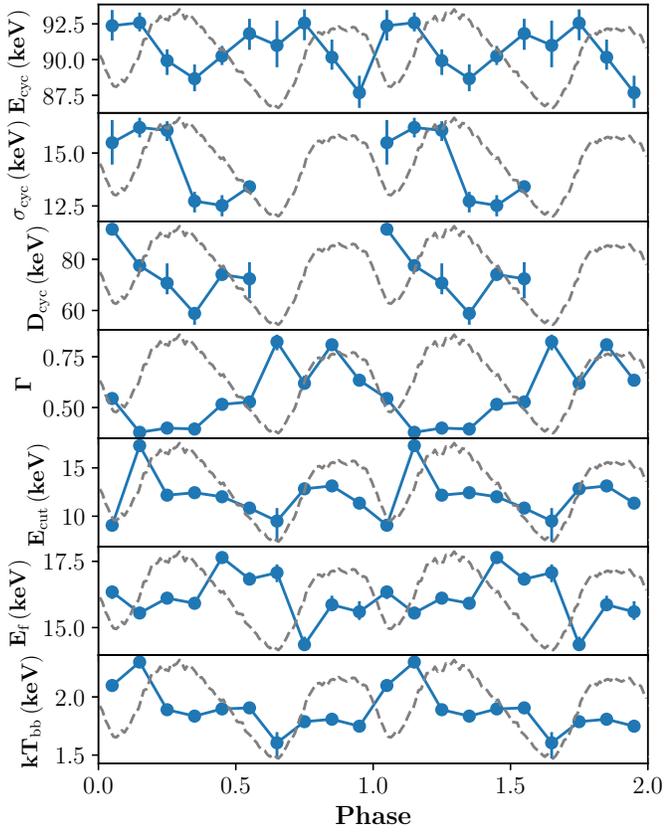}
	\caption{{Phase-resolved spectral best fit parameters, obtained with the HIGHECUT and GABS models, superimposed onto the pulse profile in the energy range 1-5 keV. The line centroid appears to have a minimun around the profile peaks.}
	}  
	\label{phaseResolved}
\end{figure}

\citet{Bellm2014, Yamamoto2014} performed pulse-phase resolved spectroscopy on 2012 giant outburst data, but no significance dependence of the line parameters on the pulse phase was measured. For the \textit{Insight-HXMT} observation discussed here, as it can be seen from Figure~\ref{phaseResolved}, the line energy appears to have a minimum at the peaks of the pulse profile, which is expected in the super-critical regime when an accretion column is present. In this case, the pulse maximum corresponds to a viewing angle, at which the largest fraction of the column, or of the neutron star's surface illuminated by the column \citep{2013ApJ...777..115P}, is visible, since in this case, the line forming region includes the largest range of the magnetic field values. The field becomes in fact weaker both with the distance from the neutron stars surface, and towards the magnetic equator. The line width is expected to increase in this case, which does not appear to be the case, at least for the main peak. So no straightforward conclusion on the the accretion regime of the source can be obtained from the phase resolved analysis.

Establishing the accretion regime of the source (sub- or super-critical accretion) requires further observational studies of the luminosity dependence of the CRSF. A final answer regarding the critical luminosity of the source, and its accretion regime, can only be obtained through additional broad-band, high statistics  observations of the source, both at lower and higher luminosity. These observations can provide significant measurements of the line parameters at different luminosity levels, obtained with the same instrument and using the same models. In addition, the monitoring of the source as a function of the luminosity can allow us to establish a break in the sign of the correlation between the line energy dependence and the luminosity, similar to that reported by \cite{Doroshenko2017,Vybornov2018}. This letter shows that new \textit{Insight-HXMT} observations can indeed solve the puzzle of the accretion regime of GRO~J1008-57. It also shows the capability of the mission to discover CRSFs at energies higher than what accessible until now, and perhaps closer to the magnetic critical field.

\acknowledgments
This work made use of the data from the \textit{Insight-HXMT} mission, a project funded by China National Space Administration (CNSA) and the Chinese Academy of Sciences (CAS). The \textit{Insight-HXMT} team gratefully acknowledges the support from the National Program on Key Research and Development Project (Grant No. 2016YFA0400800) from the Minister of Science and Technology of China (MOST) and the Strategic Priority Research Program of the Chinese Academy of Sciences (Grant
No. XDB23040400). The authors thank supports from the National Natural Science Foundation of China under Grants No. 11503027, 11673023, 11733009, U1838201, U1838202, U1838104 and U1938103.

\begin{table}	
	\centering
\caption{Best-fitting parameters of GRO J1008-57 observed with \textit{Insight-HXMT} in 2017.}
	\begin{tabular}{lrrrr}
\hline
& \multicolumn{2}{c}{HIGHECUT} & \multicolumn{2}{c}{NPEX} \\
& GABS & CYCLABS & GABS & CYCLABS\\
\hline
 $E_{\rm cyc}$ (keV) & $90.32_{-0.28}^{+0.32}$ & $84.34_{-0.43}^{+0.39}$  &  $87.35_{-0.37}^{+0.34}$  &  $83.00_{-0.63}^{+0.91}$ \\
$\sigma_{\rm cyc}$/$W_{f}$ (keV) & $14.57_{-0.11}^{+0.14}$ & $15.48_{-0.17}^{+0.14}$  &  $12.42_{-0.09}^{+0.09}$  &  $14.51_{-0.82}^{+0.98}$ \\
 $D_{\rm cyc}$ (keV)/ $D_{\rm f}$& $65.01_{-1.58}^{+1.81}$ & $2.12_{-0.02}^{+0.02}$  &  $51.34_{-0.70}^{+0.92}$  &  $1.90_{-0.11}^{+0.11}$ \\
$kT_{\rm bb}$(keV) & $1.91_{-0.01}^{+0.01}$ & $1.95_{-0.02}^{+0.02}$  &  $1.95_{-0.01}^{+0.02}$  &  $2.01_{-0.02}^{+0.02}$ \\
 $\Gamma$/ $\Gamma_1$& $0.51_{-0.01}^{+0.01}$ & $0.55_{-0.01}^{+0.02}$  &  $0.17_{-0.01}^{+0.01}$  &  $0.21_{-0.01}^{+0.01}$ \\
$E_{\rm f}$(keV) & $16.11_{-0.04}^{+0.04}$ & $17.19_{-0.09}^{+0.14}$  &  $10.96_{-0.04}^{+0.05}$  &  $11.34_{-0.09}^{+0.05}$ \\
$E_{\rm cut}$(keV) & $10.87_{-0.11}^{+0.07}$ & $12.10_{-0.29}^{+0.66}$  &  --  &  -- \\
$C_{\rm ME/LE}$ & $1.00_{-0.01}^{+0.01}$ & $0.99_{-0.01}^{+0.01}$  &  $0.99_{-0.01}^{+0.01}$  &  $0.98_{-0.01}^{+0.01}$ \\
$C_{\rm HE/LE}$& $0.99_{-0.01}^{+0.01}$ & $0.98_{-0.01}^{+0.01}$  &  $0.99_{-0.01}^{+0.01}$  &  $0.97_{-0.01}^{+0.01}$ \\
Reduced-$\chi^2$ (dof) & 1.11 (1375) & 1.14 (1375) & 1.12 (1377) & 1.16 (1377) \\
$\rm Flux_{1-100}$ ($\rm erg/s/{cm}^2$) &  \multicolumn{4}{c}{$1.43 \times 10^{-8}$}\\

\hline
	\end{tabular}
	\label{pars}
\end{table}	



\bibliography{ref}
\bibliographystyle{aasjournal}



\end{document}